\documentclass[12pt]{article}
\usepackage{color}
\usepackage{amssymb,amsmath,amsfonts}
\usepackage{epsfig}
\pdfoutput=1
\include{graphicsx}

\parskip=\medskipamount            % space between paragraphs (LaTeX)
\def\7#1#2{\mathop{\null#2}\limits^{#1}}        % puts #1 atop #2

\def\beee{\begin{equation}}
\def\eeee{\end{equation}}

\def\dels {{\nabla \hspace{-8.7pt} \slash}\;}

\begin{document}
\bibliographystyle{unsrt}

\begin{center}

\textbf{STUDY OF THE VACUUM MATRIX ELEMENT\\ OF PRODUCTS OF PARAFIELDS}\\
[5mm]
%\end{center}
O.W. Greenberg\footnote{email address: owgreen@umd.edu.}\\
{\it Center for Fundamental Physics\\
Department of Physics\\
University of Maryland\\
College Park, MD~~20742-4111}\\
and\\
A.K. Mishra\footnote{email address: mishra@imsc.res.in}\\
{\it Institute of Mathematical Sciences\\
Chennai 600 113, India}\\
University of Maryland Preprint PP-09-049\\
~\\
arXiv:0909.4069\\
\end{center}

\begin{abstract}

We study the vacuum matrix elements of products of parafields
using graphical and combinatorial methods.

\end{abstract}

\section{Introduction}
    We can calculate matrix elements of parafields using either the trilinear
    commutation relations of Green, or using the Green ansatz,
    also due to Green in the same paper~\cite{gre}.
We recently gave a path integral
    quantization of parastatistics~\cite{gremis}, primarily to show that
    parastatistics can be quantized using path integrals. We found that,
in both the canonical and path integral formalisms, the
    Green ansatz is convenient
for calculating quantities in parafield theories. This is in contrast
    with the trilinear commutation relations that are cumbersome to use for
    calculations.

As far as we know, there is no general calculation of the
    vacuum matrix element of an arbitrary product of parafields in the
    literature. Such a calculation will be useful in applications of
    parastatistics, such as a calculation of the partition function and its
    connection with the counting of partitions in number theory~\cite{tra}. In
    addition this calculation presents combinatorial problems
    that are interesting in their own right. In this paper we present the explicit
    calculation of the vacuum matrix element of products of $N \leq 6$
  scalar parabose fields (and spinor parafermi fields) of any order $p$.
We also give examples of
terms in the vacuum matrix elements of $N=8$ and $N=12$ parafields. In addition
we give methods that
are helpful to calculate the vacuum matrix elements to arbitrary order.

\section{The vacuum matrix element of a product of $N$
   parabose fields of order $p$}

   The matrix element of an arbitrary product of free bose fields is the sum
   over all products of contractions, each of which is a two-point function
with coefficient one. This is the ``factor pairing
   theorem.''~\cite{dys,wic}

For the corresponding case with parabose fields,
   not all coefficients are one; indeed the main difficulty in giving a
   general result for parabose fields is to calculate the coefficient of each
   term. These coefficients are positive or negative integers or zero.

When we work in $x$-space we use the Pauli-Jordan commutator function which is odd in $x$,
\beee
i\Delta(x)=\frac{1}{(2 \pi)^3} \int d^4k \epsilon(k^0) \delta(k^2-m^2) exp(-i k \cdot x),
\eeee
and
\beee
i\Delta(x) = \Delta^{(+)}(x) - \Delta^{(-)}(x)
\eeee
where
\beee
\Delta^{(\pm)}(x)=\int d^4k \theta(\pm k^0) \delta(k^2-m^2) exp(-i k \cdot x).
\eeee
For completeness we also define the even function of $x$
\beee
\Delta^{(1)}(x)=\Delta^{(+)}(x) + \Delta^{(-)}(x).
\eeee
The analogous formulas for free spinor parafermi fields are
\beee
iS(x)=(i\dels{_x} +m) i\Delta(x),
%iS(x)=(i\dels{{\nabla \hspace{-8.7pt}\slash}\_x} +m) i\Delta(x),
\eeee
\beee
iS(x) = S^{(+)}(x) - S^{(-)}(x)
\eeee
where
\beee
S^{(\pm)}(x)=(i\dels{_x} +m) i\Delta^{(\pm)}(x).
\eeee
\beee
S^{(1)}(x)=S^{(+)}(x) + S^{(-)}(x).
\eeee

   The formula that we gave in our previous paper,
   \beee
   \phi^{(\alpha)}(x_1) \phi^{(\beta)} (x_2) =   (2 \delta_{\alpha \beta}
   -1 ) \phi^{(\beta)} (x_2) \phi^{(\alpha)} (x_1) + \delta_{\alpha \beta}
    i \Delta (x_1 - x_2)
   \eeee
   is very useful for this calculation. In the present context,
\beee
(2 \delta_{\alpha_i \alpha_j} - 1 )^2=4 \delta_{\alpha_i \alpha_j}
-4 \delta_{\alpha_i \alpha_j} + 1=1.                     \label{id}
\eeee
This is obvious without calculation, since
\beee
2 \delta_{\alpha_i \alpha_j} - 1= 1, i=j, \textnormal{and} -1, i \neq j;
\eeee
i.e., this expression has the value $\pm 1$ and
thus its square is $1$.
  This simple identity plays an important role.
It guarantees that the result is independent of the
  details of the procedure used to evaluate the matrix element and also is
  useful in evaluating the sums over the $\alpha$'s. (Another way to write this
expression is $2 \delta_{\alpha_i \alpha_j}-1=(-1)^{\delta_{\alpha \beta}+1}$.)
Here the $\{ \phi^{(\alpha)} \}$
   fields are the Green components  associated with
   the parabose fields $\{\Phi \}$.
The analogous formula for parafermi fields is
\beee
   \psi^{(\alpha)}(x_1) \bar{\psi}^{(\beta)} (x_2) = (1-2 \delta_{\alpha \beta})
    \bar{\psi}^{(\beta)} (x_2) \psi^{(\alpha)} (x_1) + \delta_{\alpha \beta}
    i S (x_1 - x_2)
   \eeee
We choose the following systematic way
   to evaluate a vacuum matrix element, which we illustrate for the four-point
   function,\footnote{For completeness we record the result for the two-point functions,
\beee
<0|\Phi(x_1) \Phi (x_2) |0>  = p \Delta^{(+)}(x_2-x_1)     \nonumber
\eeee
\beee
<0|\Psi(x_1) \bar{\Psi} (x_2) |0>  = p S^{(+)}(x_2-x_1)    \nonumber
\eeee}
   \begin{eqnarray}
\lefteqn{ \langle 0| \Phi (x_1) \Phi (x_2) \Phi(x_3) \Phi (x_4) |0 \rangle =} \nonumber  \\
   & & \sum_{\alpha's}
    \langle 0|  \phi^{(\alpha_1)} (x_1) \phi^{(\alpha_2)} (x_2)
   \phi^{(\alpha_3)}(x_3) \phi^{(\alpha_4)} (x_4) |0  \rangle   \label{four}
\end{eqnarray}
Let $\phi^{(\alpha)} ~
    (~ \equiv  ~ \phi^{(\alpha)( + ) } + \phi^{(\alpha )( -)}) $
    act to the left on the vacuum, so that the positive
   frequency part, i.e. the creation part $ \phi^{ (\alpha)(-)}$,
   annihilates the vacuum on the left.
   Then in Eq. (1) the $i\Delta$ function will be replaced by the
   $\Delta^{(+)}$ function. Now move $\phi^{(\alpha_1)(+)}$ to the right until
   it annihilates the vacuum on the right. This field will contract with each
   of the three fields on its right, giving a $\delta_{\alpha_1 \alpha_j}$
   factor multiplying a matrix element with a product of two fields plus a
   term in which $\phi$ has moved one transposition to the right multiplied by
   a $(2 \delta -1)$ factor. The remaining two-point functions can now be
   evaluated in a similar manner, except that the second type of
term just described with the $(2 \delta -1)$ factor
   will be absent. Where a factor
$(2 \delta_{\alpha_i \alpha_j} -1)^2$
occurs, we replace it by $1$ according to Eq.(\ref{id}).
For our illustrative case of the four-point function, this procedure,
which we call the first step in our general discussion below, yields
\begin{eqnarray}
\lefteqn{ \langle 0| \Phi (x_1) \Phi (x_2) \Phi(x_3) \Phi (x_4) |0 \rangle =} \nonumber\\
& &\sum_{\alpha's}\delta_{\alpha_1 \alpha_2} \Delta^{(+)}(x_1-x_2)
 \delta_{\alpha_3 \alpha_4} \Delta^{(+)}(x_3-x_4) +  \nonumber  \\
& &(2\delta_{\alpha_1 \alpha_2}-1)\delta_{\alpha_1 \alpha_3}
\Delta^{(+)}(x_1-x_3) \delta_{\alpha_2 \alpha_4}\Delta^{(+)}(x_2-x_4)+ \nonumber  \\
& &(2\delta_{\alpha_1 \alpha_2}-1)(2\delta_{\alpha_1 \alpha_3}-1)
\delta_{\alpha_1 \alpha_4}\Delta^{(+)}(x_1-x_4)\delta_{\alpha_2 \alpha_3}
\Delta^{(+)}(x_2-x_3).
\end{eqnarray}
For the second step of the calculation of this four-point function,
each term has two different
factors $\delta_{\alpha_i \alpha_j}$ that we use to do two of the
sums over the $\alpha$'s. By renaming the dummy $\alpha$'s, if necessary,
we reduce the result to a sum over the first two $\alpha$'s. The result
is
\begin{eqnarray}
\lefteqn{ \langle 0| \Phi (x_1) \Phi (x_2) \Phi(x_3) \Phi (x_4) |0 \rangle =} \nonumber\\
& &\sum_{\alpha_1 \alpha_2}[\Delta^{(+)}(x_1-x_2) \Delta^{(+)}(x_3-x_4) + \nonumber  \\
& &(2\delta_{\alpha_1 \alpha_2}-1) \Delta^{(+)}(x_1-x_3) \Delta^{(+)}(x_2-x_4)+ \nonumber  \\
& &(2\delta_{\alpha_1 \alpha_2}-1)(2\delta_{\alpha_1 \alpha_2}-1)
\Delta^{(+)}(x_1-x_4)\Delta^{(+)}(x_2-x_3)],
\end{eqnarray}
where we used the identity Eq.(\ref{id}) in the third term. In the third and final
step, we do the sums over the two remaining $\alpha$'s. This gives the
result
%\newpage
\begin{eqnarray}
\lefteqn{ \langle 0| \Phi (x_1) \Phi (x_2) \Phi(x_3) \Phi (x_4) |0 \rangle =} \nonumber\\
& & p^2 \Delta^{(+)}(x_1-x_2) \Delta^{(+)}(x_3-x_4) +
p(2-p) \Delta^{(+)}(x_1-x_3) \Delta^{(+)}(x_2-x_4)+   \nonumber    \\
& & p^2 \Delta^{(+)}(x_1-x_4) \Delta^{(+)}(x_2-x_3).
\end{eqnarray}

For charged spinor parafermi fields the calculation is similar, except
we keep only contractions between and field $\psi$
and its conjugate $\bar{\psi}$.

In the general case the first step of this procedure reduces
  the matrix element with $N=2n$ fields
   to a sum of $N-1$ terms, each of which will have a product of $N-2$ fields.
   The entire procedure yields $(N-1)!!$ terms, each of
  which has the product of $n$ two-point functions $\Delta^{(+)}$
  multiplied by the product of $n$
  $\delta$'s and a product of $(2 \delta -1)$ factors. These terms will be
  summed over the $N$ $\alpha$'s from $1$ to $p$ for each $\alpha$.

We represent each term by a graph containing  $N$ points
and $n$ directed lines
or ``links'' below the points representing the contractions.
We can arrive at these graphs directly by forming
all possible $(N - 1)!!$ distinct combinations of $n$-pairs involving
$N=2n$ integers $1$ to $N$. In the graphs, these integers are represented by $N$ points,
numbered in ascending order.
Subsequently, we connect the pair of points representing the
integers in a pair, by a line in the graph. As each combination has $n$ pairs, $n$ such
lines are required. We choose the line (or link) to go from the
lower numbered point to a higher numbered one.

Next we provide an algebraic correspondence for these graphs.
We start with a given graph and
associate a factor
$\delta_{\alpha_i \alpha_j}  \Delta^{(+)}(x_{\alpha_i}-x_{\alpha_j})$
with the line linking the points $i$ and $j$  .
We repeat this process for each of the $n$ links, and form the product
of all factors $\delta_{\alpha_k \alpha_l}  \Delta^{(+)}(x_{\alpha_k}-x_{\alpha_l})$.
This resulting term has to be multiplied by a product of $\{(2 \delta -1) \}$
factors. The number of crossings of the links corresponds to the number of
$(2 \delta -1)$ factors in the product. (All of the links must lie below (or above)
the line containing the $N$ points.) To arrive at their  specific values,  start with the
line originating from the point $1$, and terminating, say, at point $t$. If this
line crosses other lines originating from points $t_1, t_2,... t_q$, all lying
to the left of the  point $t$, we get a product of $\{(2\delta - 1)\}$ factors, namely,
$\{(2\delta_{\alpha_1 \alpha_j} - 1), t_1 \le j \le t_q \}$.
If the line originating from point $1$ does not cross any other line, then no
$(2\delta_{\alpha_1 \alpha_j} - 1)$ factor is introduced. Repeat this process for
the lines originating from points $2, 3, ...,n$, and for each new
crossing encountered in this process, introduce a $(2\delta - 1)$ factor.
Finally, take the product of all these $\{(2\delta -1)\}$ factors, multiply this
quantity by the product of $n$-pairs of
$\delta_{\alpha_i \alpha_j}  \Delta^{(+)}(x_{\alpha_i}-x_{\alpha_j})$ factors obtained earlier,
and sum the resulting expression from 1 to $p$ over $\alpha_1, \alpha_2, ..., \alpha_N$.

As an illustration, we consider here the N = 6 case.
Generating all the  five $0$-crossing, six $1$-crossing, three $2$-crossings and
the single $3$-crossings graphs  for $N$ =$6$,  and substituting the associated
coefficients derived  in the earlier paragraph, the  six-point function is
   \begin{eqnarray}
\lefteqn{\langle 0| \Phi (x_1) \Phi (x_2) \Phi(x_3) \Phi (x_4)
 \Phi (x_5) \Phi (x_6) | 0\rangle   } \nonumber  \\
& =  & p^3 \{
 \Delta^{(+)}(x_1-x_2) \Delta^{(+)}(x_3-x_4)   \Delta^{(+)}(x_5-x_6) +
\Delta^{(+)}(x_1-x_2) \Delta^{(+)}(x_3-x_6)   \Delta^{(+)}(x_4-x_5)  \nonumber \\
&  & +\Delta^{(+)}(x_1-x_4) \Delta^{(+)}(x_2-x_3)   \Delta^{(+)}(x_5-x_6)
+ \Delta^{(+)}(x_1-x_6) \Delta^{(+)}(x_2-x_3)   \Delta^{(+)}(x_4-x_5)  \nonumber \\
&  & + \Delta^{(+)}(x_1-x_6) \Delta^{(+)}(x_2-x_5)   \Delta^{(+)}(x_3-x_4) \} + \nonumber \\
& & p^2 (2-p)   \nonumber \\
& \times & \{\Delta^{(+)}(x_1-x_6) \Delta^{(+)}(x_2-x_4)   \Delta^{(+)}(x_3-x_5)
+      \Delta^{(+)}(x_1-x_5) \Delta^{(+)}(x_2-x_3)   \Delta^{(+)}(x_4-x_6) \nonumber \\
&   &+ \Delta^{(+)}(x_1-x_3) \Delta^{(+)}(x_2-x_6)   \Delta^{(+)}(x_4-x_5)
+  \Delta^{(+)}(x_1-x_3) \Delta^{(+)}(x_2-x_4)   \Delta^{(+)}(x_5-x_6)  \nonumber \\
&   & + \Delta^{(+)}(x_1-x_2) \Delta^{(+)}(x_3-x_5)   \Delta^{(+)}(x_4-x_6)
+  \Delta^{(+)}(x_1-x_5) \Delta^{(+)}(x_2-x_6)   \Delta^{(+)}(x_3-x_4) \}   \nonumber \\
& &+ p (2-p)^2   \nonumber \\
& \times & \{\Delta^{(+)}(x_1-x_5) \Delta^{(+)}(x_2-x_4)   \Delta^{(+)}(x_3-x_6)
+      \Delta^{(+)}(x_1-x_4) \Delta^{(+)}(x_2-x_6)   \Delta^{(+)}(x_3-x_5)  \nonumber \\
&  &+ \Delta^{(+)}(x_1-x_3) \Delta^{(+)}(x_2-x_5)   \Delta^{(+)}(x_4-x_6) \} \nonumber \\
&  &- p(p^2 - 6p + 4)
 \{\Delta^{(+)}(x_1-x_4) \Delta^{(+)}(x_2-x_5)   \Delta^{(+)}(x_3-x_6) \} \nonumber \\
\end{eqnarray}

Next, we give the following 4 cases corresponding to
all possible $5$-crossings graphs for an $8$-point function as illustrations.
The algebraic expressions for graph  (1), graph (2), graph  (3) and graph  (4),
drawn in figures (1-4),  are
\begin{eqnarray}
& &\sum_{\alpha_1,\alpha_2,...,\alpha_8}
(2\delta_{\alpha_1, \alpha_2}-1)(2\delta_{\alpha_1 \alpha_3}-1) (2\delta_{\alpha_1 \alpha_4}-1)
(2\delta_{\alpha_2 \alpha_3}-1)(2\delta_{\alpha_2 \alpha_4}-1)  \times  \nonumber \\
& &\delta_{\alpha_1 \alpha_5}\Delta^{(+)}(x_1-x_5)\delta_{\alpha_2 \alpha_6}  \Delta^{(+)}(x_2-x_6) \times  \nonumber \\
& & \delta_{\alpha_3, \alpha_8}\Delta^{(+)}(x_3-x_8)\delta_{\alpha_4 \alpha_7}  \Delta^{(+)}(x_4-x_7),
\end{eqnarray}

\begin{eqnarray}
& &\sum_{\alpha_1,\alpha_2,...,\alpha_8}
[(2\delta_{\alpha_1 \alpha_2}-1)(2\delta_{\alpha_1 \alpha_3}-1) (2\delta_{\alpha_2, \alpha_3}-1)
(2\delta_{\alpha_2 \alpha_5}-1)(2\delta_{\alpha_3 \alpha_5}-1) \times \nonumber \\
& &\delta_{\alpha_1 \alpha_4}\Delta^{(+)}(x_1-x_4)\delta_{\alpha_2 \alpha_6}  \Delta^{(+)}(x_2-x_6) \times  \nonumber \\
& & \delta_{\alpha_3 \alpha_7}\Delta^{(+)}(x_3-x_7)\delta_{\alpha_5 \alpha_8}  \Delta^{(+)}(x_5-x_8)],
\end{eqnarray}
\begin{eqnarray}
& &\sum_{\alpha_1,\alpha_2,...,\alpha_8}
[(2\delta_{\alpha_1 \alpha_2}-1)(2\delta_{\alpha_1 \alpha_3}-1) (2\delta_{\alpha_1 \alpha_4}-1)
(2\delta_{\alpha_2 \alpha_4}-1)(2\delta_{\alpha_3 \alpha_4}-1) \times \nonumber \\
& &\delta_{\alpha_1 \alpha_5}\Delta^{(+)}(x_1-x_5)\delta_{\alpha_2 \alpha_7}  \Delta^{(+)}(x_2-x_7) \times  \nonumber \\
& & \delta_{\alpha_3 \alpha_6}\Delta^{(+)}(x_3-x_6)\delta_{\alpha_4 \alpha_8}  \Delta^{(+)}(x_4-x_8)],
\end{eqnarray}
and
\begin{eqnarray}
& &\sum_{\alpha_1,\alpha_2,...,\alpha_8}
[(2\delta_{\alpha_1 \alpha_3}-1)(2\delta_{\alpha_1 \alpha_4}-1) (2\delta_{\alpha_2 \alpha_3}-1)
(2\delta_{\alpha_2 \alpha_4}-1)(2\delta_{\alpha_3 \alpha_4}-1) \times \nonumber \\
& &\delta_{\alpha_1 \alpha_6}\Delta^{(+)}(x_1-x_6)\delta_{\alpha_2 \alpha_5}  \Delta^{(+)}(x_2-x_5) \times  \nonumber \\
& & \delta_{\alpha_3 \alpha_7}\Delta^{(+)}(x_3-x_7)\delta_{\alpha_4 \alpha_8}  \Delta^{(+)}(x_4-x_8)],
\end{eqnarray}
respectively.

%\begin{figure}
%\includegraphics{figure1.eps}
%\includegraphics[scale=1]{figure1.eps}
%\end{figure}
Since there are $n$ $\delta$-functions in each one of the $(N -1)!!$ terms, in the second
step we reduce the
expressions to sums over $n$ $\alpha$'s.
%The next question is what is the maximum number of links or crossings
%for a $N$-point function.
We now define a
graph with $n$ vertices and as many links between vertices
as there are crossings in the diagrams with $N$ points discussed above.
%vertices discussed above.
A link connects a pair of vertices if lines
starting on each of the pair intersect. The most
``saturated'' graph with the maximum number, $n(n-1)/2$
links corresponds to a simplex in $n$-dimensional space. This also implies
that the coefficients contain at most a product of [$n(n-1)/2$]
$(2\delta - 1)$ factors.
In the final step
we do the sums over the remaining $n$ $\alpha$'s. The total number of summations
to be carried out is $(N - 1)!!$. Obviously for a graph with
$m$ crossings ($ 1 \le m \le n(n-1)/2 $), the summand is a product
of $m$ $(2\delta - 1)$ terms. When $m$ = 0, summand is unity, and the coefficient
equals $p^n$. For $m$  = 1, the summand is
$(2\delta_{\alpha_i,\alpha_j} - 1)$, and the associated  coefficient
takes the value $p^{n - 1} (2 - p)$. When $m$ = 2, all possible values of
the summand, through relabeling the dummy indices $\{\alpha_i\}$,
get reduced to two forms, viz, $(2\delta_{\alpha_1,\alpha_2} - 1)
(2\delta_{\alpha_1,\alpha_3} - 1)$,  and
$(2\delta_{\alpha_1,\alpha_2} - 1)   (2\delta_{\alpha_3,\alpha_4} - 1)$.
%and $(2\delta_{\alpha_1,\alpha_2} - 1) (2\delta_{\alpha_5,\alpha_6} - 1)$; the last one
%arising only when $N \ge 4$.
The coefficients corresponding to
both the  summands are $p^{n-2}(2-p)^2$.  For $m$ = 3, $N$ = 6, only one
graph with three crossings exists; having the summand as
$(2\delta_{\alpha_1,\alpha_2} - 1) (2\delta_{\alpha_1,\alpha_3} - 1)
(2\delta_{\alpha_2,\alpha_3} - 1)$, and the coefficient is  $-p(p^2-6p+4)$.
In general, when $m$ takes the maximum value $n(n - 1)/2$, the
 is $\prod_{i < j;  1 \le i \le n-1; 2 \le  j \le n}
(2\delta_{\alpha_i,\alpha_j} - 1 )$, and there is only one graph corresponding
to this most saturated case.

As a step towards finding a general method to evaluate the  coefficients, we can
simplify the problem of  $(N - 1)!!$ summations for $N$ number of fields
by enumerating the
number of graphs  $T_{Nm}$ having  $m$ numbers of crossings.
As noted earlier, $ 0 \le m \le n(n-1)/2 $.
In the following table,
these values are listed for  $ 2 \le N \le 10$. The $^*$ in a box implies that
the $m$ has reached its maximum permissible value.

\newpage
\begin{center}
TABLE I
\end{center}

\begin{center}
\begin{tabular}{|c|c|c|c|c|c|c|c|c|c|c|c|c|} \hline
%& & & & & & & & & & &  \\
%\hline
$m$ & 0 & 1 & 2 & 3 & 4 & 5 & 6 & 7 & 8 & 9 & 10 &  \begin{tabular}{c}$\sum_m  T_{Nm}$ \\
$ = (N - 1)!! $   \end{tabular}
\\
\hline
& & & & & & & & & & & & \\
$T_{2m}$ & $1^*$  & & & & & & & & & & & 1
\\
\hline
& & & & & & & & & & & & \\
$T_{4m}$ & 2 & $1^*$  & & & & & & & & & & 3
\\
\hline
& & & & & & & & & & & & \\
$T_{6m}$ & 5 & 6 & 3 & $1^*$  & & & & & & & & 15
\\
\hline
& & & & & & & & & & & & \\
$T_{8m}$ & 14 & 28 & 28 & 20 & 10 & 4 & $1^*$ & & & & & 105
\\
\hline
& & & & & & & & & & & & \\
$T_{10m}$ & 42 & 120 & 180 & 195 & 165 & 117 & 70 & 35 & 15 & 5 & $1^*$ & 945
\\
\hline
\end{tabular}
\end{center}

Now we provide a general method for carrying out summation over $\alpha$'s.
We express the sum from 1 to $p$ over each of the $n$ $\alpha$'s as a sum over
the partitions of $n$.
In our example of the four-point function, we have reduced the calculation
of, for example, the numerical coefficient of the middle term in Eq. (6), to the sum
\beee
\sum_{\alpha_1,\alpha_2=1}^p(2 \delta_{\alpha_1,\alpha_2}-1)
\eeee
Instead of summing each $\alpha$ from $1$ to $p$, we can break up the sums
into a sum over single values of each $\alpha$, for example
\beee
\sum_{\alpha_i=1}^p=\sum_{\alpha_i=1} + \sum_{\alpha_i=2}+ \cdots +
\sum_{\alpha_i=p}
\eeee
When we carry out multiple sums, we have to take account whether there
is a $(2 \delta -1)$ factor connecting a pair of $\alpha$'s and which sets
of $\alpha$'s match and which don't match. If there is no $(2 \delta -1)$
factor connecting a pair of $\alpha$'s there is a factor $1$; if there is
a $(2 \delta -1)$
factor connecting a pair of $\alpha$'s there is a factor $-1$.
Next, we select any specific term,
and sum over the set of, say, $h$ $\alpha$'s,
that do not appear in the summand. Obviously
summation over these $h$ $\alpha$'s leads to a multiplicative
factor of $p^h$ in the coefficients. Now we are left with the
summation over the remaining $(n -h) \equiv f $ $\alpha$'s. To make further
progress, we begin with the case of $h$ = 0 corresponding to
the maximum number of crossings. Here not only all $\alpha$'s
appear in the summand, but each one appears for the equal  number
of times. This symmetric distribution of $\alpha$'s in each summand
leads to considerable simplification. Our further discussion applies
only to this ``maximally saturated'' case, although the use of partitions
with restrictions may apply generally.

For this symmetric case we group the numbers from
$1$ to $n$ into groups that are equal and unequal. This
leads to unrestricted partitions of the number $n$.
Let the $p(n)$ partitions of $n$ be labeled $\lambda^{(1)},
\lambda^{(2)},\cdots,\lambda^{(p(n))}$. For a general
$\lambda^{(s)}$ let
\beee
\lambda^{(s)} = ( \lambda^{(s)}_1...\lambda^{(s)}_1\lambda^{(s)}_2...
\lambda^{(s)}_2...\lambda^{(s)}_i...\lambda^{(s)}_i...\lambda^{(s)}_l...\lambda^{(s)}_l)
\eeee
where $m_i^{(s)}$ is the number of times the part $\lambda^{(s)}_i$ of the partition
$\lambda^{(s)}$ occurs,
$k^{(s)}=\sum_{i=1}^{l^{(s)}}m_i^{(s)}$ is the number of parts of the partition
$\lambda^{(s)}$,
$l^{(s)}$ is the number of distinct parts, and the weight is
$|\lambda^{(s)}|=\sum_{(i=1)}^{l^{(s)}}m_i^{(s)} \lambda_i^{(s)} = n$. Then
\beee
\sum_{\alpha_1 = 1}{^p} \cdots \sum_{\alpha_n = 1}{^p}f(\alpha_1, \cdots,\alpha_n)
 = \sum_{s = 1}^{p(n)} \frac{\displaystyle n!}
 {\displaystyle \prod_{i=1}^{l(s)}[(\lambda_i^{(s)} !)^{m_i^{(s)}}m_i^{(s)}!]}
\frac {\displaystyle p!}
{\displaystyle (p - k^{(s)})!} f(\alpha_1, \cdots,\alpha_n)|_{\lambda^{(s)}_i}                         \label{general}
\eeee
where $p^n$ is the total number of   `n-tuples'
$(\alpha_1, \alpha_2, ..., \alpha_n)$
when each $\alpha$ varies from 1 to p.
In our case, the $f(\alpha_1, \cdots,\alpha_n)|_{\lambda^{(s)}}$
is the product of the factors $(2\delta -1)$
associated with the crossings discussed above. This product is either 1 or -1 for
a given partition. For the case of no crossings,
whose $n$-vertex graphs
consist of $n$ disconnected points, $f(\alpha_1, \cdots,\alpha_n)|_{\lambda^{(s)}}=1$
and we find
\beee
p^n=\sum_{s = 1}^{p(n)} \frac{\displaystyle n!}
 {\displaystyle \prod_{i=1}^{l(s)}[(\lambda_i^{(s)} !)^{m_i^{(s)}}m_i^{(s)}!]}
\frac {\displaystyle p!}
{\displaystyle (p - k^{(s)})!}                                      \label{p^n}
\eeee
as we expect.
The summand for any  `n-tuple' of
$(\alpha_1, \alpha_2,.., \alpha_n)$
always reduces to either $1$ or $-1$. The partitions introduced above determine this
sign. As noted above
$(2\delta_{\alpha_i,\alpha_j} -1 ) = 1 $  if $\alpha_i$ = $\alpha_j$
and $-1$ otherwise. If in any given partition $\lambda^{(s)}$, $\alpha_i$  is not equal
to $\alpha_j$ for $A^{(s)}$ number of times, the summand would be $(-1)^{A^{(s)}}$. This is true
for all the  possible groupings of $\alpha$'s  according to the
partition $\lambda^{(s)}_i$. This is a consequence of the fact that in the most saturated case
the $\alpha$'s have
a symmetric distribution in the summand. From Eq. (\ref{general}), we get the total number of
configurations
$E^{(s)}$ associated with
the $\lambda^{(s)}$ as
\beee
E^{(s)} = X^{(s)}_n
\frac {\displaystyle p!}
{\displaystyle (p - k^{(s)})!}  \ ; \ X^{(s)}_n =                              \label{p^m}
 \frac{\displaystyle n!}
 {\displaystyle \prod_{i=1}^{l(s)}[(\lambda_i^{(s)} !)^{m_i^{(s)}}m_i^{(s)}!]}
\eeee

The summand contains a product of $n(n-1)/2 ~$ $(2\delta{\alpha_i,\alpha_j} -1)$ terms,
and in the partition $\lambda^{(s)}$, the number of times for which $\alpha_i$ equals $\alpha_j$ is
\beee
\bar{A}^{(s)} =      \sum_{i=1}^l m_i^{(s)}
\frac {\displaystyle \lambda^{(s)}_i (\lambda^{(s)}_i - 1 )}
 {\displaystyle 2} \ \ ;  \ \   \lambda^{(s)}_i \ge 2.
\eeee
Consequently,
\beee
A^{(s)} =
\frac {\displaystyle n (n - 1 )}
 {\displaystyle 2} - \bar{A}^{(s)}
%- \sum_{i=1}^l m_i^{(s)}
%\frac {\displaystyle \lambda^{(s)}_i (\lambda^{(s)}_i - 1 )}; \lambda^{(s)} \ge 2.
% {\displaystyle 2}
\eeee
and the coefficient
\beee
\sum_{\alpha_1,..., \alpha_n} \prod_{i < j;  2 \leq \{i+1, j\} \leq n}
(2\delta_{\alpha_i,\alpha_j} - 1 )
= \sum_{s=1}^{p(n)} (-1)^{A^{(s)}} E^{(s)}
\eeee

As an illustration, we evaluate the coefficient $C^{12}_{15}$
for the most saturated graph corresponding to $N$ = $12$, viz.,
\beee
C^{12}_{15}\ \  =
\sum_{\alpha_1,..., \alpha_6}
 \prod_{i < j;  2 \le \{i+1, j\} \le 6}
(2\delta_{\alpha_i,\alpha_j} - 1 )
\eeee
It contains a product of $15$
$(2\delta - 1)$ factors or a sum of $2^{15}$ terms. The relevant parameters for calculating
$C^{12}_{15}$ are given in the TABLE II. Using Eq. (32) and Table II, we get
 %  \begin{eqnarray}
\beee
C^{12}_{15}   =   -p(p-2)^2 (p^3-26p^2+152p-128).
%\end{eqnarray}
\eeee
For bosons, $p =1$, and  $C^{12}_{15}$ should, and does, reduce to $1$.

\newpage
\begin{center}
TABLE II
\end{center}
\begin{center}
\begin{tabular}{|c|c|c|c|c|c|c|c|} \hline % c|c|c|c|c|} \hline
$s$ & $\lambda^{(s)}$ & $\lambda^{(s)}_1/m^{(s)}_1$ &$ \lambda^{(s)}_2/m^{(s)}_2$ & $\lambda^{(s)}_3/m^{(s)}_3$ &$ k^{(s)}$
& $E^{(s)}$ & $A^{(s)}$
\\
\hline
& & & & & & &    \\
$1$ & [6]  & 6/1 &- &- &1  & 6!p!/\{6!(p-1)!\} &0
\\
\hline
& & & & & & &  \\
$2$ & [5,1] &5/1  &1/1 &- & 2& 6!p!/\{5!(p-2)!\} & 5
\\
\hline
& & & & & & &  \\
$3$ & [4,2] &4/1  &2/1 &- & 2& 6!p!/\{4!2!(p-2)!\} & 8
\\
\hline
& & & & & & &  \\
$4$ & [4,1,1] &4/1  &1/2 &- & 3& 6!p!/\{4!2!(p-3)!\} & 9
\\
\hline
& & & & & & &  \\
$5$ & [3,3] &3/2  &- &- & 2& 6!p!/\{3!$^2$2!(p-2)!\} & 9
\\
\hline
& & & & & & &  \\
$6$ & [3,2,1] &3/1  &2/1 &1/1 & 3& 6!p!/\{3!2!(p-3)!\} & 11
\\
\hline
& & & & & & &  \\
$7$ & [3,1,1,1] &3/1  &1/3 &- & 4& 6!p!/\{3!$^2$(p-4)!\} & 12
\\
\hline
& & & & & & &  \\
$8$ & [2,2,2] &2/3  &- &- & 3& 6!p!/\{2!$^3$3!(p-3)!\} & 12
\\
\hline
& & & & & & &  \\
$9$ & [2,2,1,1] &2/2  &1/2 &- & 4& 6!p!/\{2!$^4$(p-4)!\} & 13
\\
\hline
& & & & & & &  \\
$10$ & [2,1,1,1,1] &2/1  &1/4 &- & 5& 6!p!/\{2!4!(p-5)!\} & 14
\\
\hline
& & & & & & &  \\
$11$ & [1,1,1,1,1,1] &1/6  &- &- & 6& 6!p!/\{1!$^6$6!(p-6)!\} & 15
\\
\hline
\end{tabular}
\end{center}

 \section {Summary and Conclusions}

 We used Green's ansatz to evaluate the
 vacuum matrix element of an arbitrary product of parabose fields of order
 $p$. We gave a graphical method to reduce an $N=2n$ point
 function to a sum over products of $n$ $2$-point functions. We expressed the coefficients
 of each these $n$ $2$-point functions as a
 sum over $n$ $\alpha_i$'s ($1 \le \alpha_i \le p$).
 The corresponding graph uniquely determines both the $n$ $2$-point functions and the
 coefficients that occur in the sum.

 Subsequently, we gave a combinatorial method, based on the partitions
 of the number $n$, to determine the coefficients. We illustrated this method
by the explicit determination of the coefficient for most
 saturated graph corresponding to the\\ $N$ = 12 case which involves a sum of $2^{15}$ terms.

\section {Figure Captions}

Figure 1. Case (1) of a 5-crossings graph for an 8-point function. The algebraic
expression associated with this graph is given in Eq. (20)

\noindent
Figure 2. Case (2) of a 5-crossings graph for an 8-point function. The algebraic
expression associated with this graph is given in Eq. (21)

\noindent
Figure 3. Case (3) of a of 5-crossings graph for an 8-point function. The
associated algebraic expression for this graph is given in Eq. (22)

\noindent
Figure 4. Case(4) of a of 5-crossings graph for an 8-point function. The
associated algebraic expression for this graph is given in Eq. (23)

\section {Acknowledgements:}
%We happy to thank for valuable discussions.
This work was supported in part by the National Science Foundation,
Grant No. PHY-0140301 and by the Department of Science and
Technology of India Grant. No. DST/INT/US (NSF-RP086).

\begin{figure}
\begin{center}
    \includegraphics[scale=.75]{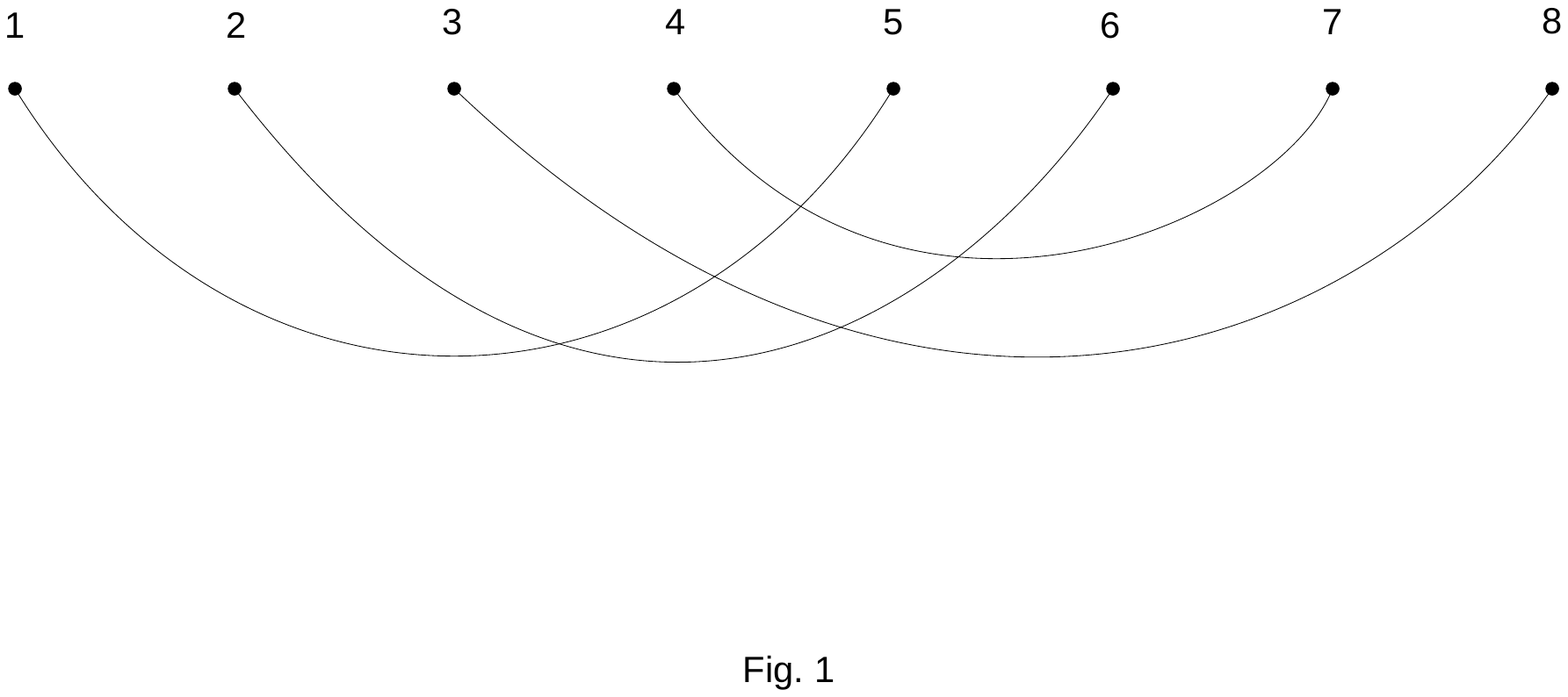}
\end{center}
%\caption{Case (1) of a 5-crossings graph for an 8-point function. The %algebraic
%expression associated with this graph is given in Eq. (20)}
\end{figure}
\begin{figure}
\begin{center}
    \includegraphics[scale=.75]{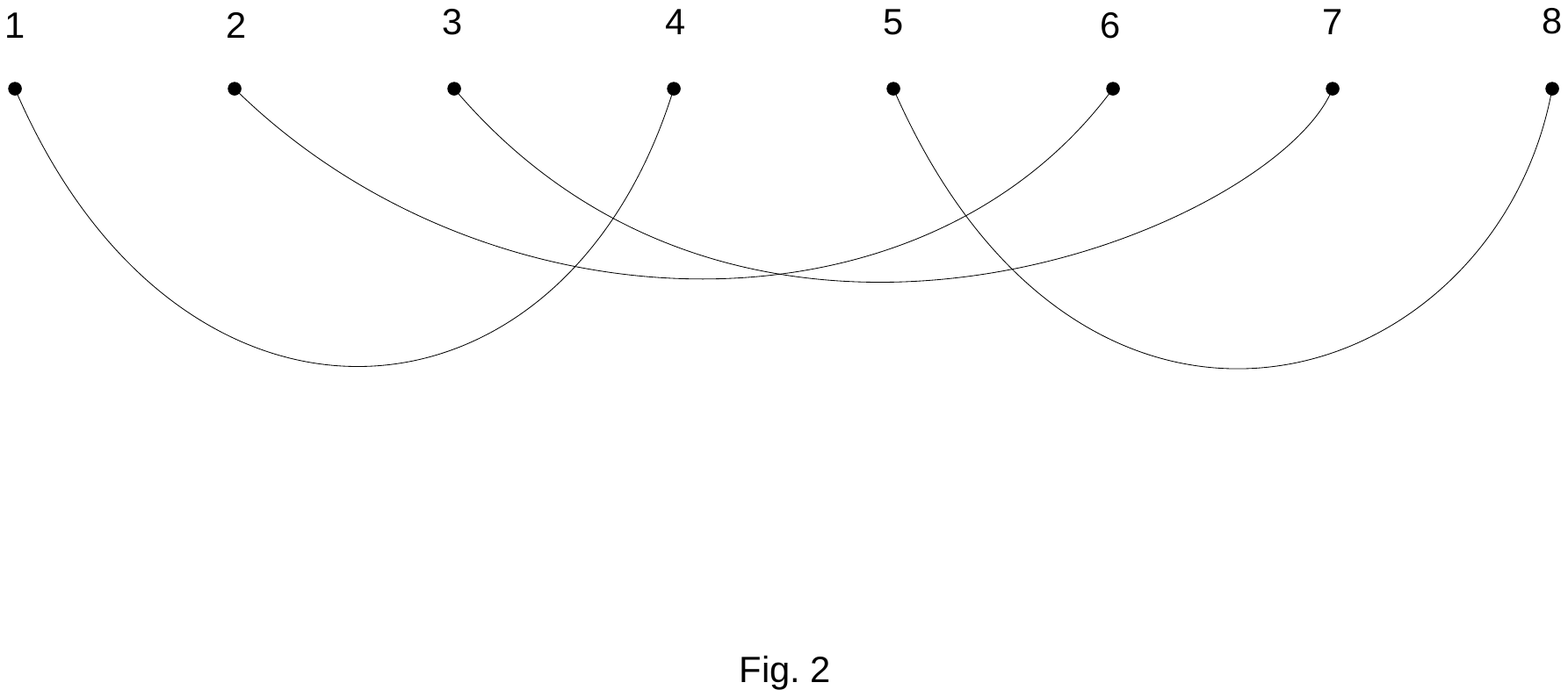}
\end{center}
%\caption{Case (2) of a 5-crossings graph for an 8-point function. The %algebraic
%expression associated with this graph is given in Eq. (21)}
\end{figure}
\begin{figure}
\begin{center}
    \includegraphics[scale=.75]{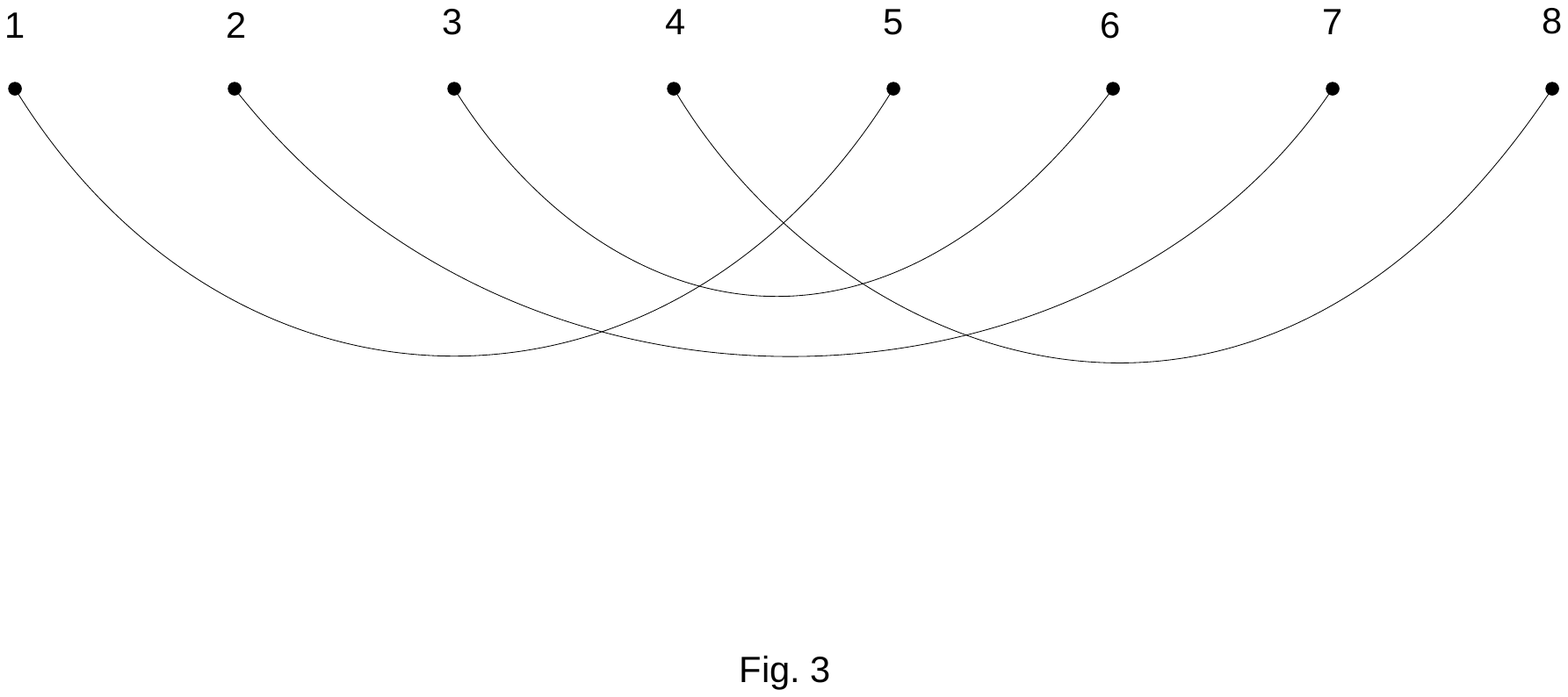}
\end{center}
%\caption{Case (3) of a of 5-crossings graph for an 8-point function. The
%associated algebraic expression for this graph is given in Eq. (22)}
\end{figure}
\begin{figure}
\begin{center}
    \includegraphics[scale=.75]{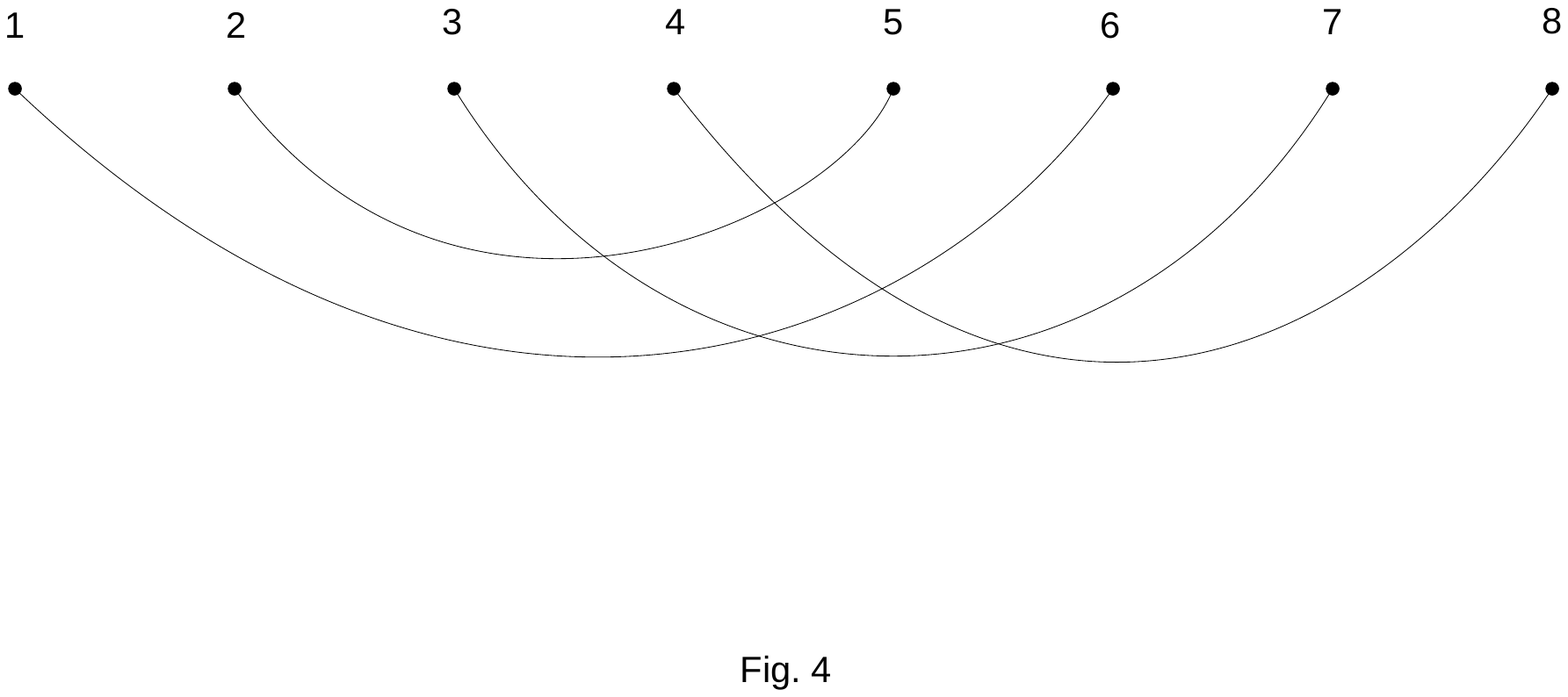}
\end{center}
%\caption{Case (4) of a of 5-crossings graph for an 8-point function. The
%associated algebraic expression for  this graph is given in Eq. (23)}
\end{figure}
\end{document}